\documentclass[sigconf]{acmart}
 \usepackage{amsmath,amssymb,amsfonts,bm}
\usepackage{graphicx}
\usepackage{textcomp}
\usepackage{xcolor}
\usepackage{float}
\usepackage{array}
\usepackage{tabularx}
\usepackage{array}
\usepackage{balance}
\usepackage{marvosym}
\usepackage{multirow}
\usepackage{array}
\usepackage{enumitem}
\usepackage{booktabs}
\usepackage{algpseudocode}
\usepackage[ruled]{algorithm2e} 
\usepackage{xcolor}
\usepackage{pifont}
\usepackage{url}
\usepackage{algpseudocode}
\usepackage{subfig}
\usepackage[ruled]{algorithm2e}
\begin{document}
\copyrightyear{2023}
\acmYear{2023}
\setcopyright{acmlicensed}\acmConference[CIKM '23]{Proceedings of the 32nd ACM International Conference on Information and Knowledge Management}{October 21--25, 2023}{Birmingham, United Kingdom}
\acmBooktitle{Proceedings of the 32nd ACM International Conference on Information and Knowledge Management (CIKM '23), October 21--25, 2023, Birmingham, United Kingdom}
\acmPrice{15.00}
\acmDOI{10.1145/3583780.3615088}
\acmISBN{979-8-4007-0124-5/23/10}
%%
%% The "title" command has an optional parameter,
%% allowing the author to define a "short title" to be used in page headers.
\title{Towards Communication-Efficient Model Updating for On-Device Session-Based Recommendation}

\author{Xin Xia}
\affiliation{%
	\institution{The University of Queensland}
	\city{Brisbane}
	\country{Australia}}
\email{x.xia@uq.edu.au}

\author{Junliang Yu}
\affiliation{%
	\institution{The University of Queensland}	
	\city{Brisbane}
	\country{Australia}}
\email{jl.yu@uq.edu.au}

\author{Guandong Xu}
\affiliation{%
	\institution{University of Technology Sydney}
		\city{Sydney}
	\country{Australia}}
\email{Guandong.Xu@uts.edu.au}

\author{Hongzhi Yin}
\authornote{Corresponding author.}
\affiliation{%
	\institution{The University of Queensland}
	\city{Brisbane}
	\country{Australia}}
\email{h.yin1@uq.edu.au}
\fancyhead{}

%%
%% The abstract is a short summary of the work to be presented in the
%% article.
\begin{abstract}
	On-device recommender systems recently have garnered increasing attention due to their advantages of providing prompt response and securing privacy. To stay current with evolving user interests, cloud-based recommender systems are periodically updated with new interaction data. However, on-device models struggle to retrain themselves because of limited onboard computing resources. As a solution, we consider the scenario where the model retraining occurs on the server side and then the updated parameters are transferred to edge devices via network communication. While this eliminates the need for local retraining, it incurs a regular transfer of parameters that significantly taxes network bandwidth. To mitigate this issue, we develop an efficient approach based on compositional codes to compress the model update. This approach ensures the on-device model is updated flexibly with minimal additional parameters whilst utilizing previous knowledge. The extensive experiments conducted on multiple session-based recommendation models with distinctive architectures demonstrate that the on-device model can achieve comparable accuracy to the retrained server-side counterpart through transferring an update 60x smaller in size. The codes are available at \url{https://github.com/xiaxin1998/ODUpdate}.
	\end{abstract}
%extremely high compression rate
%%
%% The code below is generated by the tool at http://dl.acm.org/ccs.cfm.
%% Please copy and paste the code instead of the example below.
%%
% \begin{CCSXML}
% <ccs2012>
%  <concept>
%   <concept_id>10010520.10010553.10010562</concept_id>
%   <concept_desc>Computer systems organization~Embedded systems</concept_desc>
%   <concept_significance>500</concept_significance>
%  </concept>
%  <concept>
%   <concept_id>10010520.10010575.10010755</concept_id>
%   <concept_desc>Computer systems organization~Redundancy</concept_desc>
%   <concept_significance>300</concept_significance>
%  </concept>
%  <concept>
%   <concept_id>10010520.10010553.10010554</concept_id>
%   <concept_desc>Computer systems organization~Robotics</concept_desc>
%   <concept_significance>100</concept_significance>
%  </concept>
%  <concept>
%   <concept_id>10003033.10003083.10003095</concept_id>
%   <concept_desc>Networks~Network reliability</concept_desc>
%   <concept_significance>100</concept_significance>
%  </concept>
% </ccs2012>
% \end{CCSXML}

% \ccsdesc[500]{Computer systems organization~Embedded systems}
% \ccsdesc[300]{Computer systems organization~Redundancy}
% \ccsdesc{Computer systems organization~Robotics}
% \ccsdesc[100]{Networks~Network reliability}

%%
%% Keywords. The author(s) should pick words that accurately describe
%% the work being presented. Separate the keywords with commas.
% \begin{CCSXML}
% <ccs2012>
%    <concept>
%        <concept_id>10002951.10003317.10003347.10003350</concept_id>
%        <concept_desc>Information systems~Recommender systems</concept_desc>
%        <concept_significance>500</concept_significance>
%        </concept>
%  </ccs2012>
% \end{CCSXML}

% \ccsdesc[500]{Information systems~Recommender systems}


\maketitle
\begin{figure}[t]
    \centering
    \includegraphics[width=0.5\textwidth]{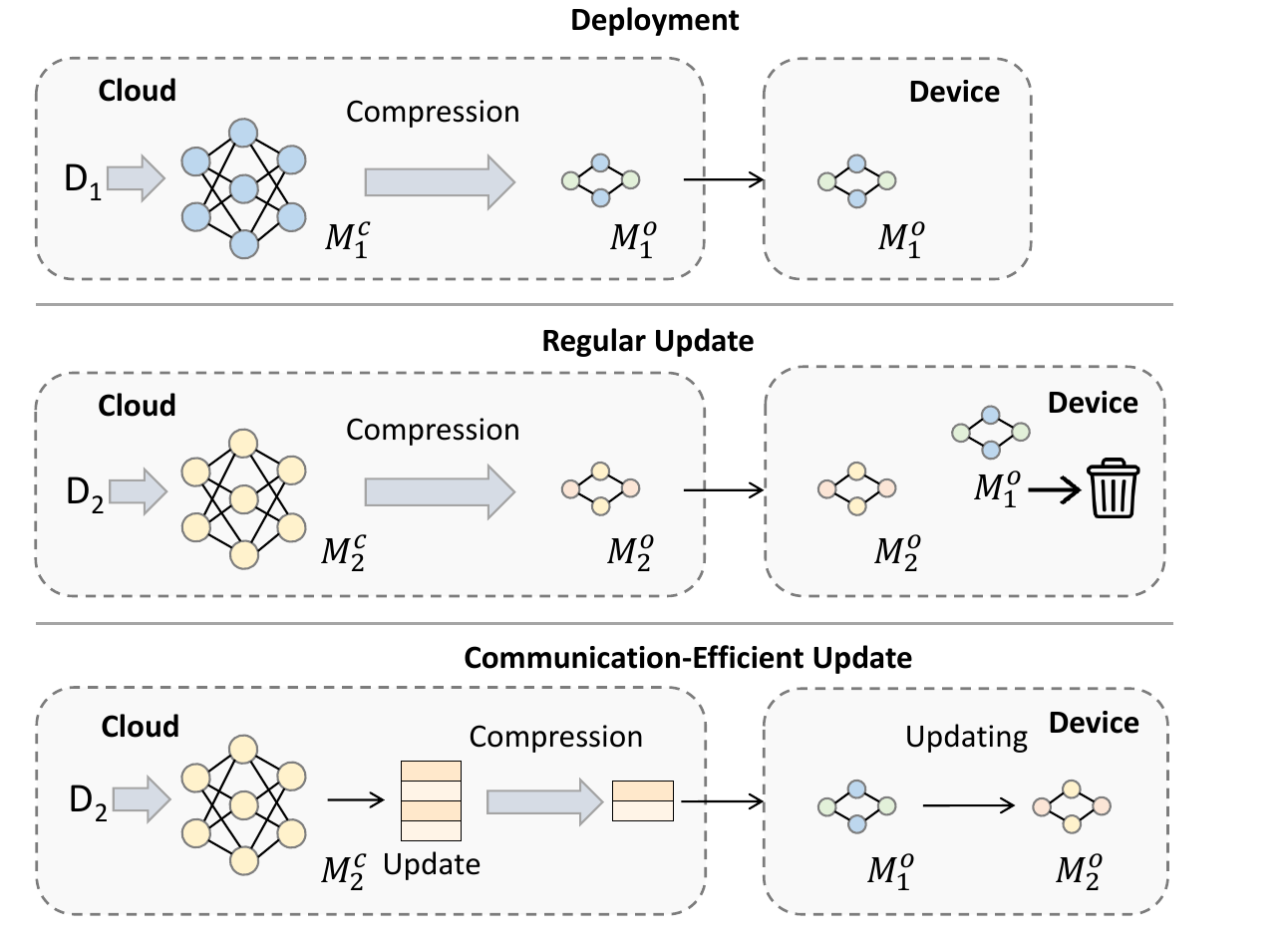}
    \caption{Deployment and update of on-device recommendation models (D represents Data and M denotes Model). }
    \label{figure.1}
    \vspace{-10pt}
\end{figure}
\section{Introduction}
Deep learning has bestowed recommender systems with exceptional capabilities in the past decade, demonstrating unparalleled outcomes across a diverse range of applications \cite{zhang2019deep, yu2023self, nguyen2017argument, hung2017computing, wang2018streaming}. Nonetheless, the maintenance and training expenses of neural-architecture-based recommendation models have escalated substantially due to their growing size and compute-intensive nature, necessitating a significant amount of computing power, storage, and memory while incurring a considerable carbon footprint \cite{himeur2021survey, xia2021self, yu2022graph, junliang2021socially, xia2021co, yu2023self, yu2023xsimgcl}. Furthermore, centralized models on the cloud often need personal data in order to generate immediate recommendations, which poses a risk of privacy leakage \cite{jeckmans2013privacy}. To counteract these limitations, on-device machine learning \cite{dhar2021survey}, as an energy-saving paradigm, has garnered increasing attention in the field of recommender systems. This emerging machine learning paradigm focuses on the design and implementation of lightweight models on resource-constrained devices, providing users with low-latency and secure services. It has been proven to be an effective solution  to address the limitations of current cloud-based recommender systems, with various on-device recommendation models demonstrating promising performance \cite{han2021deeprec, wang2020next, ochiai2019real, changmai2019device, xia2022device}. \par

Currently, the predominant focus of the research on on-device recommender systems \cite{xia2022device, xia2023efficient} revolves around the model compression techniques aimed at accommodating memory-constrained devices. Since the substantial memory consumption is attributed primarily to the embedding table of items/users and their associated features rather than the model depth, various model compression techniques, such as low-rank decomposition \cite{oseledets2011tensor}, hash coding \cite{chen2021learning}, and quantization \cite{gong2014compressing}, have been integrated and refined in different recommendation models to shrink the embedding table. While the proliferation of compression techniques has provided solutions to fit recommendation models into edge devices, there has been a long-standing neglect of the critical issue of efficiently updating on-device models \cite{chen2022update}. Due to the continuous evolution of user interests, maintaining recommender systems is an ongoing task. As millions of users generate new interactions all the time, failure to retrain the model with the new data results in a performance decline over time. Although cloud-based recommender systems can leverage powerful computing resources and abundant memory to continuously retrain deep models, such an approach is not feasible for devices with limited resources. One direct solution to update deployed models is periodically transferring retrained models to the device and discarding obsolete ones, as depicted in Figure \ref{figure.1}. Nevertheless, even with significant model compression prior to transfer, frequent model updates still incur substantial network communication costs. \par

To mitigate this issue, in this paper, we propose a communication-efficient way to enable recurrent model update with only a tiny subset of parameters. To this end, our recommendation model is first compressed using compositional codes \cite{shu2017compressing} prior to the deployment. Specifically, each item in the model is represented by a short code used to index continuous vectors from several codebooks which replace the item embedding table. When the server accumulates a certain volume of new data, the cloud model is retrained to acquire a new code matrix and updated codebooks. As the item embeddings encode the essential information, the newly introduced parameters within the code matrix and codebooks are deemed to reflect the most crucial modifications to the model, referred to as the model update. In order to minimize the communication cost, inspired by \cite{chen2022update}, we propose to further compress the model update to efficiently maintain the on-device model. 

Two novel approaches for compressing model updates that integrate traditional data structures, specifically stack and queue, are put forward in this paper. These approaches selectively transfer and update a subset of new parameters, using the principles of Last-In-First-Out (LIFO) and First-In-First-Out (FIFO), respectively. In the queue-based approach, we add the newest parameters to one end of the queue, while an equal number of old parameters are removed from the other end. Analogously, in the stack-based approach, we pop out some old parameters from the stack before appending the new parameters. By combining previous parameters with new parameters, our approaches optimize the performance of on-device models at a tiny communication cost. However, these approaches still require the pre-definition of a fixed compressed ratio prior to retraining, which can lead to sub-optimal performance when the ratio is not adequately chosen. To achieve self-adaptiveness, we introduce the Maximum Mean Discrepancy (MMD) metric \cite{gretton2012kernel} to measure the data distribution shift, which enables flexible adjustment of the update size. A larger MMD indicates a substantial data shift between item embeddings, necessitating the consideration of a larger update size for stable recommendation quality. Finally, to validate the generalizability of the proposed approaches, we apply them to multiple on-device session-based recommendation models with distinctive architectures and obtain encouraging results.
% The simplest one is selecting a part of vectors in codebooks to transfer and then integrates with the parameters from the original codebooks under $D_1$. This method utilizes existing knowledge and integrates with new update to ensure recommendation quality. To maximize the utilization of existing knowledge, we propose another method where traditional structure, stack is integrated. We keep all the old parameters and then add a subset of new parameters for both code matrix and codebooks. The new parameter follows the rule of stack, where the previous addition parameter is discarded when the next new parameter arrives. The third method integrates the rule of queue, which means that the old parameters which have the same size of the new will be discarded from the other side of the queue when new parameters come in. 
% When doing experiments, there is a trade-off between the update size and the recommendation improvement. Our method is represented in Figure \ref{figure.2}.
% We apply our method in session-based recommendation scenario. Through extensive experimental results, we can achieve the goal of updating on-device recommendation model with comparable accuracy to the retrained server-side model through transferring a update that is 10x smaller in size than the original update.

Overall, our contributions are summarized as follows:

\begin{itemize}[leftmargin=*]
    \item We are the first to investigate the issue of communication-efficient update for on-device recommendation models by introducing the concept of update compression. 
    \item We develop two efficient model-agnostic approaches for on-device model update, which leverage traditional data structures (queue and stack), to enable progressive and self-adaptive replacement of model parameters. These approaches ensure that on-device models remain up-to-date while utilizing previous knowledge effectively.
    \item Extensive experiments conducted on two benchmarks and various session-based recommendation models demonstrate that the on-device model can achieve comparable accuracy to the retrained server-side counterpart through transferring an update 60x smaller in size.
\end{itemize}

\section{Related Work}
\subsection{On-Device Recommendation}
On-device recommendation is gaining popularity recently due to its ability to perform low latency inferences and enhance privacy protection. This field of research \cite{ changmai2019device, dhar2021survey, han2021deeprec, ochiai2019real, chen2021learning} mainly employs model compression techniques, such as pruning \cite{han2015deep, srinivas2015data}, quantization \cite{gong2014compressing}, and low-rank factorization \cite{novikov2015tensorizing, oseledets2011tensor}, to reduce the size of traditional models to make them fit edge devices, achieving acceptable performance while utilizing minimal model sizes. Specifically, WIME \textit{et al.} \cite{changmai2019device} proposes a method for generating on-device recommendations where a neighborhood searching method followed by a sequence matching algorithm was used to ensure efficient on-device prediction. Wang \textit{et al.} \cite{wang2020next} were the first to propose using tensor-train factorization to compress embedding table for next POI recommendation model. TT-Rec \cite{yin2021tt}, also applies tensor-train decomposition to compress the embedding table, and improves performance by introducing a cache to store important item's embeddings. OD-Rec \cite{xia2022device} then enhances tensor-train decomposition by replacing the multiplications with semi-tensor decomposition to make it more flexible and efficient and applies it into session-based scenarios. Chen \cite{chen2021learning} proposed the use of elastic item embeddings to make lightweight models adaptable to various devices under varying memory constraints for on-device recommendation. DeepRec \cite{han2021deeprec}, on the other hand, utilizes model pruning and embedding sparsification techniques for model training and fine-tunes the model using local user data, achieving comparable performance in sequential recommendation tasks with a reduction of 10x in terms of model parameters. These models manage to deploy compact recommendation models on edge devices but they do not consider the problem of updating on-device recommendation models. 

\subsection{Model Update Compression}
Update compression was first explored in Federated Learning \cite{konevcny2016federated} where the goal of update compression is to minimize the communication overhead while preserving the accuracy of the model updates and the training data is required to be available in clients.
The approaches to do model update in federated learning can be divided into two types, i.e. low-rank update, gradient-based methods. In \cite{konevcny2016federated}, low-rank update factorizes the gradient tensor into two matrices where only one matrix is optimized in training and another matrix is generated randomly and compressed as random seed to be transmitted with the optimized matrix. Gradient sparsification methods \cite{aji2017sparse,seide20141} involve the hard-thresholding of smaller gradients and the transmission of only the larger, deemed important ones. Qu \textit{et al.} \cite{qu2022deep} proposed gradient-based deep partial updating paradigm where only the parameters make the largest contributions are updated in each round to achieve a similar performance with full updating. Gradient quantization approach \cite{strom2015scalable} reduces the precision of parameters to have a trade-off between accuracy and update size. Gradient-based methods are different from low-rank updates because the update size is reduced during training and iterations. Chen \textit{et al.} \cite{chen2022update} proposed to compress the update for DNNs on edge based on low-rank approximation which is similar to our idea, but it is not for recommendation scenario and the SVD technique used in the paper is not efficient. In this paper, we are the first to do update compression in on-device recommendation and we assume the non-sensitive training data is available in the cloud. 

% In recommender systems, PeterRec \cite{yuan2020parameter} proposes a parameter-efficient way to transfer user profiles to accomplish model fine-tuning but it is not for on-device recommendation.
\subsection{Compositional Encoding}
To overcome the high storage requirements of one-hot encoding, more efficient encoding systems such as Huffman Code \cite{han2015deep,huffman1952method} and Hash functions \cite{tito2017hash, chen2015compressing} have been developed. However, these encoding systems often suffer from limited representation capability, which can result in decreased accuracy. As a solution, subsequent works \cite{chen2018learning, shu2017compressing} have investigated addictive quantization for source coding, leading to the development compositional codes. In recommender systems, compositional code has been adopted to compress the model size \cite{liu2019compositional, shi2020compositional, kang2020learning}. Liu \textit{et al.} \cite{liu2019compositional} proposed to use compositional codes to represented items/users with a set of binary vectors associated with a sparse weight vector which encodes the importances of each binary codes, achieving great efficiency improvement. DHE \cite{kang2020learning} encodes the feature value to a unique identifier vector with multiple hashing functions and transformations, and then applies a DNN to convert the identifier vector to an embedding, which is very close to compositional encoding and has comparable accuracy with smaller model size against one-hot encoding. Shi \textit{et al.} \cite{shi2020compositional}  exploited complementary partitions of the category set to produce a unique embedding vector for each category without explicit definition. They achieved this by storing multiple smaller embedding tables based on each complementary partition and combining embeddings from each table. But these methods either require category or feature information or bear a high computational cost. Li \textit{et al.} \cite{li2021lightweight} utilized compositional codes to shrink item embedding table in the sequential recommendation. However, the model adopts a quotient-remainder trick to form distinct codes for items which increases the time complexity. In this paper, we provide an efficient and effective compositional coding method without using any feature or user information.

\section{Problem Setting}
In this paper, we focus on the efficient update for on-device session-based recommendation models. It is reasonable to assume that the on-device model can not retrain itself due to limited computing resources and memory storage and the newly collected interaction data is available in the server side. There exists communication bandwidth limits between the server and the resource-constrained devices.

\subsection{Session-Based Recommendation Task}
To validate the efficiency of the proposed methodology, we apply it to the scenario of on-device session-based recommendation \cite{wang2021survey}, which is very common in real-world recommender systems. Let $\mathcal{V} = \{v_{1}, v_{2}, v_{3}, ... , v_{|\mathcal{V}|}\}$ denote all items. The items within each session sequence are arranged in chronological order, and each session was generated by an anonymous user. The goal of session-based recommendation is to predict the next item that a user will interact with for the current session. We map and embed every item $v\in \mathcal{V}$ into the embedding space, so the item embedding table is denoted as $\mathbf{X}\in\mathbb{R}^{|\mathcal{V}| \times d}$. Given $\mathcal{V}$ and a session $s$, the output of a session-based recommendation model is a ranked list $y = [y_{1}, y_{2}, y_{3}, ..., y_{|\mathcal{V}|}]$ where $y_{v}$ is the predicted probability of item $v$ being recommended. The top-\textit{K} items $(1 \leq K \ll |\mathcal{V}|)$ with highest probabilities will be presented as the recommendation list.

\subsection{On-Device Model Deployment}
Let $D_1=\left\{\left(s_i, l_i\right)\right\}_{i=1}^{N_1}$ denote the dataset in the initial stage where $s_i$ represents $i$-th sequence and $l_i$ represents its label. In the initial stage, we have a large session-based recommendation model in the cloud denoted as $M_{1}^{c}(\Theta_1 \mid D_1)$ with parameters $\Theta_1$, which is fully trained with dataset $D_1$. At first, we have:
\begin{equation}
	Y_1^{c}=\underset{\Theta_1}{\arg \min }\mathcal{L}\left(f(\Theta_1, D_1)\right),
\end{equation}
where $\mathcal{L}$ is the training loss of recommendation task (e.g., cross entropy), $f$ is the session-based recommendation task and $Y_1^{c}$ is predicted probability.
To deploy $M_{1}^{c}$ on resource-constrained devices, we compress it into a compact model $M_{1}^{o}(\Theta_1\mid D_1)$ and tune it using $D_1$:
\begin{equation}
	Y_1^{o}=\underset{\theta_1}{\arg \min }\mathcal{L}\left(f(\theta_1, D_1)\right),
\end{equation}
where $Y_1^{o}$ is the predicted probability, $\theta_1$ is the parameters of the compressed model, and the compression ratio is $\Theta_1/\theta_1$. Then the compressed model $M_1^{o}$ is downloaded and deployed on devices.
\begin{figure*}[t]
	\centering
	\includegraphics[width=\textwidth]{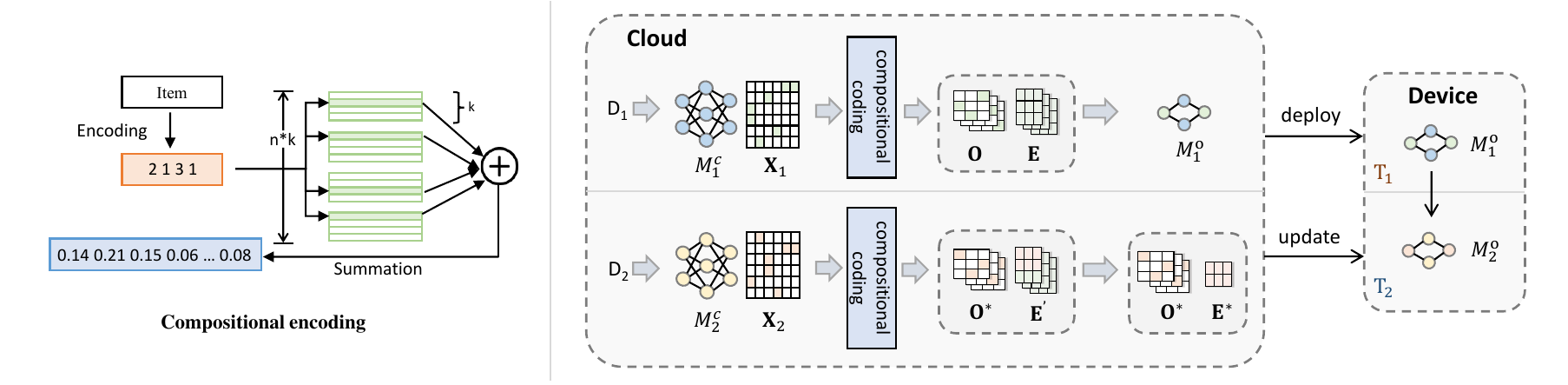}
	\caption{(Left): Illustration of compositional encoding. (Right): An overview of on-device update compression.}
	\label{figure.2}
\end{figure*}
\subsection{On-Device Model Update}
After deployment, when the cloud collects new training data $D_2=\left\{\left(s_i, l_i\right)\right\}_{i=1}^{N_2}$ and $N_2>N_1$ and $D_2 \supset D_1$, the cloud-side model $M^{c}_{1}$ will be retrained on $D_2$ to be $M^{c}_{2}(\Theta_2 \mid D_2)$. To update the on-device model, a direct approach is compressing the retrained recommendation model and then transfer it to devices. The previous model $M_{1}^{o}(\theta_1 \mid D_1)$ is discarded and then $M_{2}^{o}(\theta_2 \mid D_2)$ is set up. However, as retraining models is frequently required, the communication cost is huge for recurrent update and there exists bandwidth limits. Therefore, we propose to compress the updated parameters $\theta_2$ given $D_2$ and $M^{c}_{1}$, which is formulated as:
\begin{equation}
	\Delta= Compress\left(M_{1}^{c}\left(\Theta_1 \mid D_1\right), D_2\right),
\end{equation} 
where $\Delta$ denotes the compressed update with a tiny size. After that, $\Delta$ is transmitted to the device. At the device side, the current model can integrate $\Delta$ to reconstitute and accomplish model update, which is formulated as:

\begin{equation}
	M^{o}_2\left(\theta_2 \mid D_2\right)= Update\left(M^{o}_1\left(\theta_1 \mid D_1\right), \Delta\right).
\end{equation}

\section{Methodology}
In this section, we present a framework that seamlessly connects the model compression and update compression for on-device session-based recommender systems, as depicted in Fig. \ref{figure.2}.
\subsection{Model Compression with Compositional Coding}
For model compression, tensor-train decomposition and quantization are two commonly used techniques in related research. But they sometimes fail to provide enough flexibility in compression and struggle to achieve significant compression ratios. To deal with these issues, we introduce compositional coding \cite{shu2017compressing} to compress the item embedding table $\mathbf{X}$. Instead of representing items with continuous embedding vectors, we represent each item by a unique $n$-dimensional code consisting of $n$ discrete codewords, where $n$ is much smaller than the original embedding dimension $d$. We then learn embedding vectors for each codeword rather than each item and then simply sum up the corresponding codeword vectors to construct item embedding. Compositional codes can represent item embeddings as a combination of smaller, reusable components that allows for greater modeling flexibility and expressive power. \par

Specifically, each item $v$ is represented by a unique code $C_v = (C_v^1, C_v^2, ..., C_v^n) \in \mathcal{B}^{n}$ where each component in the code is ranged in [1, $k$] and $\mathcal{B}$ is the set of code bits with cardinality of $n$. There we learn a code allocation function $\Phi(\cdot): \mathcal{V} \rightarrow \mathcal{B}^{n}$ that maps each item with its discrete code. To generate the corresponding continuous embedding vector based on the discrete code of item $v$, we first embed $C_v$ to a sequence of embedding vectors $(\bm{E}_1^{C_v^1}, \bm{E}_2^{C_v^2}, ..., \bm{E}_n^{C_v^n})$. Here we create $n$ codebooks $\bm{E}_1, \bm{E}_2, ..., \bm{E}_n$, each containing $k$ vectors (i.e. basic vectors) with the dimension of $d$. The final embedding of item $v$ is: 
\begin{equation}
	\bm{e}_v=\sum_{i=1}^{n} \bm{E}_{i}\left(C_{v}^{i}\right),
\end{equation}
where $\bm{E}_{i}\left(C_{v}^{i}\right)$ is the $C_{v}^{i}$-th vector in codebook $\bm{E}_i$. Note that the sample space for this summation-based embedding combination is ${nk \choose n}$ which should be larger than $|\mathcal{V}|$ to provide a unique code for each item. \par

% Compositional encoding significantly reduces the model size, therefore the model performance will degrade drastically.
% To maintain model capacity in next-item recommendation, namely, given baseline embedding matrix $\mathbf{X}$, we want to find a set of codes $C$ and codebooks $\bm{E}$ that can generate embeddings with the same effectiveness as $\mathbf{X}$. We directly minimize the squared distance between the original embedding and the composite embedding:
% \begin{equation}
% 	\begin{aligned}
% 	\mathcal{L}_{mse} &=\underset{C, \bm{E}}{\operatorname{argmin}} \frac{1}{|\mathcal{V}|} \sum_{v \in \mathcal{V}}\left\|\bm{E}(C_{v})-\mathbf{X}_v\right\|^{2} \\
% 	&=\underset{C, \bm{E}}{\operatorname{argmin}} \frac{1}{|\mathcal{V}|} \sum_{v \in \mathcal{V}}\left\|\sum_{i=1}^{M} \bm{E}_{i}(C_{v}^{i})-\mathbf{X}_v\right\|^{2}.
% 	\end{aligned}
% \end{equation}

% To minimize the MSE loss means we should optimize the item-to-code mapping function $\Phi$ and the code-to-embedding composition function (Eq.(5)).
However, the representation learning is not differentiable because of the discreteness of compositional codes. In order to make it end-to-end, we consider each code $C_v$ as a concatenation of $n$ one-hot vectors, i.e. $C_v = (\bm{O}_1^v, \bm{O}_2^v, ..., \bm{O}_n^v)$, where $\forall i, \bm{O}_{i}^{v} \in[0,1]^{k}$ and $\sum_{m} \bm{O}_{i}^{v m}=1$, and $\bm{O}_{i}^{v m}$ is the $m$-th component of $\bm{O}_i^v$. Let $\bm{O}_1, \bm{O}_2, .., \bm{O}_n$ represent $n$ one-hot matrices where each matrix is with the size of $|V| \times k$ and each row in each matrix is a $k$-dimensional one-hot vector. 
Then the generation of item's embedding can be reformulated as:
\begin{equation}
	\bm{e}_v=\sum_{i=0}^{n} \bm{O}_{i}^{v}\bm{E}_{i},
\end{equation}
where $\bm{O}_i^v$ represents the one-hot vector corresponding to the code component $C_v^i$ of item $v$.
Inspired by \cite{shu2017compressing}, we then adopt Gumbel-Softmax \cite{JangGP17} to make the continuous vector $\bm{O}_i^v$ approximate the one-hot vector. The $m$-th element in $\bm{O}_i^v$ is computed as:
\begin{equation}
	\begin{aligned}
	\left(\bm{O}_{i}^{v}\right)_{m} &=\operatorname{softmax}\left(\log \boldsymbol{\alpha}_{v}^{i}+G\right)_{m} \\
	&=\frac{\exp \left(\left(\log \left(\boldsymbol{\alpha}_{v}^{i}\right)_{m}+G_{m}\right) / \tau\right)}{\sum_{m^{\prime}=1}^{k} \exp \left(\left(\log \left(\boldsymbol{\alpha}_{v}^{i}\right)_{k^{\prime}}+G_{m^{\prime}}\right) / \tau \right)},
	\end{aligned}
\end{equation}
where $G_m$ is a noise term that is sampled from the Gumbel distribution $-\log (-\log ($Uniform$(0,1))$, $\tau$ is the temperature in softmax. As $\tau\rightarrow$ 0, the softmax computation smoothly approaches the $\arg$max. (In our setting, we set $\tau$ to 0.1). And $\boldsymbol{\alpha}_{v}^{i}$ is computed by a two-layer MLP:  
\begin{equation}
	\begin{aligned}
	\boldsymbol{\alpha}_{v}^{i} &=\operatorname{softmax}\left(\operatorname{softplus}(\boldsymbol{\phi}_{i}^{\prime \top} \boldsymbol{h}_{v}+\boldsymbol{b}_{i}^{\prime})\right) \\
	\boldsymbol{h}_{v} &=\tanh \left(\boldsymbol{\phi}^{\top} \mathbf{X}_{v}+\boldsymbol{b}\right),
	\end{aligned}
\end{equation}
where $\mathbf{X}_{v}$ is the well-trained embedding of item $v$, $\boldsymbol{\phi} \in  \mathbb{R}^{d \times (nk/2)}$, $\boldsymbol{\phi}_{i}^{\prime} \in \mathbb{R}^{(nk/2) \times nk}$, $\boldsymbol{b} \in \mathbb{R}^{nk/2}$ and $\boldsymbol{b}^{\prime} \in \mathbb{R}^{nk}$. Benefiting from this, the model can circumvent the indifferentiable look-up operation and enable the code learning end-to-end in our method.
The learning of compositional codes and vectors is done in server-side. After being compressed and optimized, the model can finally fit in resource-constrained devices. To get item embeddings in bulk, we can concatenate each codebook and each one-hot code matrix together which is be formulated as:
\begin{equation}
	\mathbf{X}= [\bm{O}_1, \bm{O}_2, .., \bm{O}_n][\bm{E}_1, \bm{E}_2, .., \bm{E}_n]^{\top} = \bm{O}\bm{E},
\end{equation}
where $\bm{E} = [\bm{E}_1, \bm{E}_2, .., \bm{E}_n]^{\top}$, $\bm{E} \in \mathbb{R}^{nk\times d}$ and $\bm{O}=[\bm{O}_1, \bm{O}_2, .., \bm{O}_n]$, $\bm{O} \in \mathbb{R}^{|V|\times(nk)}$. As $\bm{O}$ is a sparse matrix that can be transformed into a 1-dimensional array containing $n|\mathcal{V}|$ elements after training, which can directly look up the codebooks (shown in Fig. {(\ref{figure.2})}: Left), the compression ratio is calculated as follows:
\begin{equation}
\label{eq: cr}
	CR = \frac{|\mathcal{V}|d}{nkd + n|\mathcal{V}|}=\frac{1}{\frac{nk}{|\mathcal{V}|}+\frac{n}{d}}\,\,\,\,\,\,\,\  (nk\ll|\mathcal{V}|,\,n\ll d),
\end{equation}
where $nkd$ stands for the size of the codebooks.

% \begin{algorithm}[t]
%     \caption{The training process of the proposed method.}
%     \LinesNumbered 
%     \label{alg:Framework}
    
%     \KwIn{session data $D_1$, $D_2$, ..., $D_5$}
%     \KwOut{Recommendation lists} 
% 	Train a session-based recommendation model $M$ with parameters $\Theta$\;     
%     Train a compact model $M_1$ on $D_1$ through Eq. (6)-(9) and Eq. (21)-(23)\;
%     Deploy $M_1$ on resource-constrained devices\;
% 	Retrain $M$ on $D_2$ and use new parameters to retrain $M_1$ to get $M_2$\;
% 	Transfer compressed update and reconstitute the on-device model through Eq. (12)-(17)\;
% 	Generate recommendation lists through the updated model\;
	
% \end{algorithm} 

\subsection{Update Compression}
% To a microscopically view, compositional codes compressed the item embedding table so that the model size can be reduced and is able to be deployed on resource-constrained devices. In other words, we can see it as a re-parameterization of the recommendation model.
% The learning of compositional codes and vectors is done in server-side. After being compressed and optimized, the model can finally fit in resource-constrained devices. To get item embeddings in bulk, it can be formulated as:
% \begin{equation}
% 	\mathbf{X}= [\bm{O}_1, \bm{O}_2, .., \bm{O}_m][\bm{E}_1, \bm{E}_2, .., \bm{E}_m]^{\top} = \bm{O}\bm{E},
% \end{equation}
% where $\bm{E} = [\bm{E}_1, \bm{E}_2, .., \bm{E}_m]^{\top}$, $\bm{E} \in \mathbb{R}^{(nk)\times d}$ and $\bm{O}=[\bm{O}_1, \bm{O}_2, .., \bm{O}_m]$, $\bm{O} \in \mathbb{R}^{|V|\times(nk)}$.
% Then when we retrain the model, its new learnable parameters can be:
% \begin{equation}
% 	\theta^{\prime}=\left(\theta \backslash\left\{\mathbf{X}\right\}\right) \cup\left\{\bm{E}, \bm{O}\right\}.
% \end{equation}
% Here $\theta$ includes other parameters in the model, but when transferring updates, we only transfer the item embeddings. 
The item embedding table in a session-based recommendation model is the most memory-intensive component and conveys the most crucial information, so our proposed method focuses on its compression and update and does not deal with other parameters, as discussed in \cite{yuan2020parameter}. When updating, the problem of replacing the embedding table can be addressed by efficiently transmitting the new one-hot matrices and codebooks with which the on-device model can reconstitute the embedding table. Our goal is to achieve comparable performance to the server model with minimal parameters to be transferred. 
%In sequential tasks, the recommendation model has the tendency to abruptly forget its previously learned knowledge when it is incrementally trained on new data, which is called Catastrophic Forgetting \cite{french1999catastrophic}. This can result in decreasing performance over time. 
To this end, we propose to integrate the model update with two traditional data structures, stack and queue and where the two approaches allow the model to be updated with new parameters whilst preserving existing knowledge. As one-hot matrices are very sparse and only contain 0 and 1 that can be further compressed to be represented with 1 bit per element, thereby consuming very limited memories, we will not compress the one-hot matrices and just directly transfer them with the compressed new codebooks. We depict the update compression in Fig. \ref{figure.3}. \par
\subsubsection{Stack-Based Update}
\begin{figure}[t]
	\centering
	\includegraphics[width=0.5\textwidth]{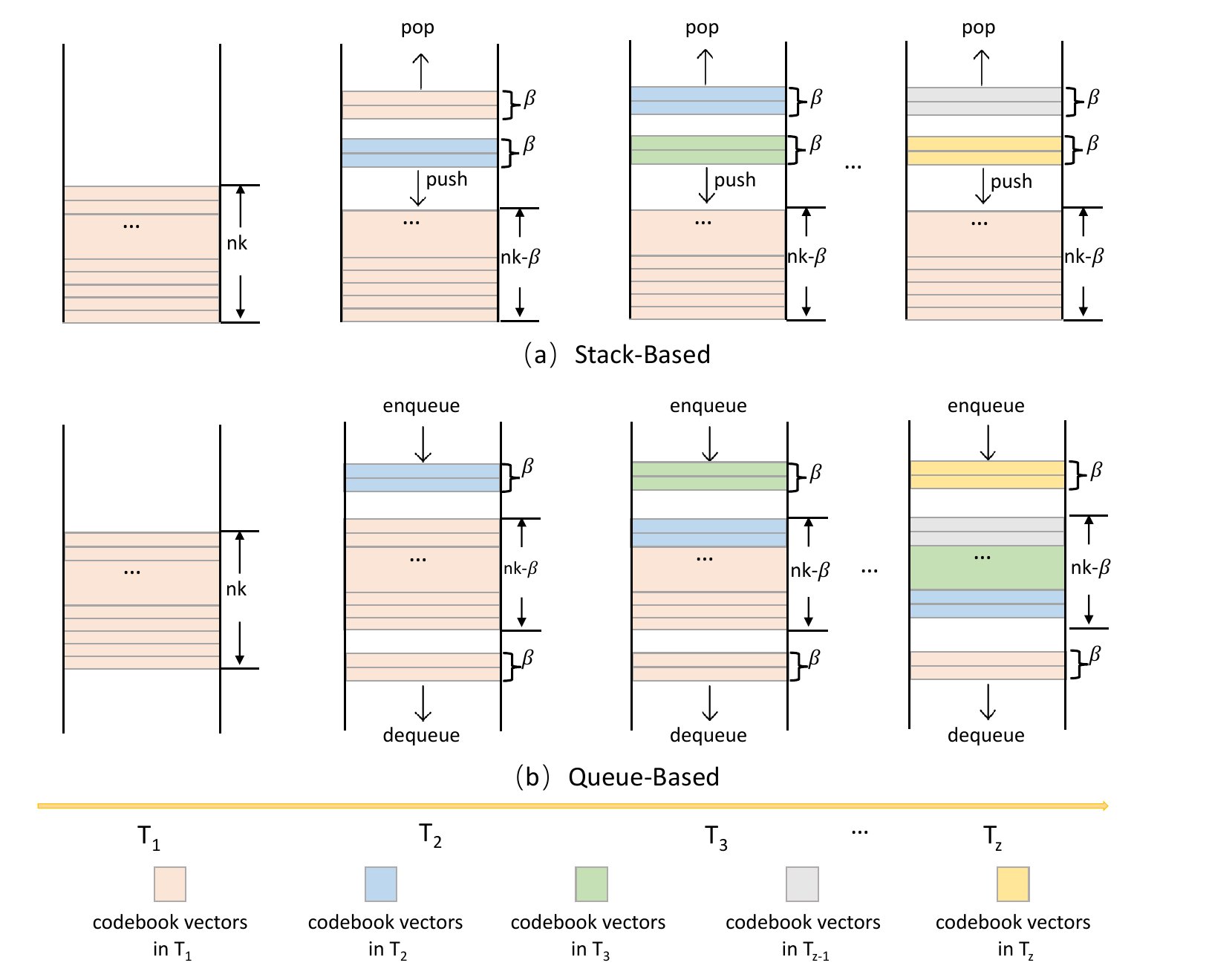}
	\caption{Two approaches for efficient model update.}
	\label{figure.3}
	\vspace{-10pt}
\end{figure}

The first approach (Fig. 3 (a)) is updating by means of a stack. Assume that we have $z$ time slices, i.e. $T_1, T_2, ..., T_z$. $T_1$ stands for the stage of model deployment. Imagine there is a stack as depicted in Fig. \ref{figure.3}, and $\bm{E}$ is pushed in as a whole at $T_1$. When the new parameters arrive in the next time slice, a subset of parameters from the top of the stack are popped out and the same number of new parameters are then pushed in. Note that the codebooks on both server-side and device side are maintained with a stack. When we update the codebooks and one-hot matrices in the cloud, we only train the parameters planned to be transferred (i.e., at the top of the stack) and keep the rest parameters frozen. Specifically, at $T_2$, if we have $n$ codebooks and we aim to achieve an update compression ratio $r$, then we only need to retrain $\beta = nk/r$ vectors where $\beta$ is an integer $\in$ [1, $nk$], and replace the top $\beta$ vectors in the stack with the new parameters. While the other parameters in the codebooks also participate in this retraining process, they are fixed to preserve the knowledge learned at $T_1$ and avoid the high-cost transfer of all the codebooks. In this way, the total parameters to be transferred are:
\begin{equation}
	\Delta = \{\bm{O}^{*}, \bm{E}^{*}\},
\end{equation}
where $\bm{O}^{*} \in \mathbb{R}^{|V|\times nk}$ denotes the new one-hot matrices, and $\bm{E}^{*} \in \mathbb{R}^{\beta \times d}$ denotes the updated parameters in the codebooks. When on-device inference is required, following the LIFO principle, the updated item embedding table is reconstituted as:
\begin{equation}
	\mathbf{X}^{\prime} = MatMul(\bm{O}^{*} , CONCAT[\bm{E}^{*}, \bm{E}[\beta:, :]]).
\end{equation}
We can tune the value of $r$ to ensure $|\Delta|\ll \left|\theta_2\right|$. In the following time slices, the on-device model is updated in the same way. \par
% \begin{figure}[t]
% 	\centering
% 	\includegraphics[width=0.5\textwidth]{figure3.pdf}
% 	\caption{Two approaches for efficient model update.}
% 	\label{figure.3}
% 	\vspace{-10pt}
% \end{figure}
% It is easy to find that the size of the two matrices are reduced while the embedding size or the model architecture is not changed. Therefore, the multiplication between them protects the size of embeddings and meanwhile reduces the parameter amount. To utilize the update to reconstitute the on-device model, we can get updated embedding as:
% However, when $\beta$ is small, the performance improvement will be very limited after update. Therefore, we propose to utilize previous knowledge $\theta_1$ which is already learned in model deployment. \par
\subsubsection{Queue-Based Update}
In the queue-based approach (Fig. 3 (b)), we also recycle most of the parameters in the codebooks. The difference is that every time when new parameters are delivered to the device, the same number of existing parameters will be popped out from the front of the queue and the new parameters are added to the back of the queue. In other words, as time passes, all the parameters learned at $T_1$ are replaced with new parameters. By contrast, in the stack-based approach, most parameters in the codebooks learned at $T_1$ are kept unless $\beta$ could be very large. Following the FIFO principle, the updated item embedding table on the device is reconstituted as:
\begin{equation}
	\mathbf{X}^{\prime} = MatMul(\bm{O}^{*},CONCAT[\bm{E}^{*}, \bm{E}[:(nk-\beta), :]]).
\end{equation}
With these two approaches, the compression ratio of an update can be calculated as:
\begin{equation}
	\frac{nkd + n|\mathcal{V}|}{\beta d + n|\mathcal{V}|}.
\end{equation} 
Compare to the strategy of transferring the original item embedding table for model update, the compression ratio of our proposed approaches is: 
\begin{equation}
\label{eq: tcr}
	CR = \frac{|\mathcal{V}|d}{nkd + n|\mathcal{V}|}\frac{nkd + n|\mathcal{V}|}{\beta d + n|\mathcal{V}|}
	= \frac{1}{\frac{\beta}{|\mathcal{V}|}+\frac{n}{d}}.
\end{equation}
Given that the queue-based approach can progressively update all the parameters, it is natural to consider it a better solution as recommender systems face the data shift problem \cite{yin2016adapting}. However, even the queue-based approach indeed demonstrates superior performance in the majority cases of our experiments, the stack-based approach can serve a distinct purpose and cater to specific needs in particular scenarios such as movie recommendation where the user's preference tends to remain stable for long. Accordingly, it is reasonable to preserve parameters learned in the early stage with a stack-based mechanism. In contrast, the queue-based approach can better serve news/short video recommendation where user preferences may change relatively fast. To ensure comprehensive coverage across a wide range of scenarios, both stack and queue-based approaches are proposed and validated in this paper.   \par
\subsection{Self-Adaptive Update Compression}
While the proposed approaches enable efficient model update, they require a manually pre-defined compressed ratio prior to retraining, which can lead to
sub-optimal performance when the ratio is not adequately chosen. Particularly, in different time slides, the degree of data shift can fluctuate and hence a self-adaptive mechanism for the selection of the compressed ratio is needed. Our idea is that we can measure the data shift by comparing the server-side item embedding tables learned at two contiguous time slices and use the shift to decide the update size flexibly and autonomously. Inspired by \cite{gretton2012kernel, liu2020learning}, we adopt Maximum Mean Discrepancy (MMD) to measure the data shift between the item embedding tables. Based on the definition of MMD, we calculate the MMD value for two embedding tables $\mathbf{X}^{(t)}$ and $\mathbf{X}^{(t+1)}$:
\begin{equation}
\label{eq: MMD}
	\begin{aligned}
		\operatorname{MMD}^{2}[\mathcal{F}, \mathbf{X}^{(t)}, \mathbf{X}^{(t+1)}] &= \frac{1}{n_1^2} \sum_{i, j=1}^{n_1} \mathcal{K} \left(x^{(t)}_{i}, x^{(t)}_{j}\right)  -\\ \frac{2}{n_1 n_2} \sum_{i, j=1}^{n_1, n_2} \mathcal{K} \left(x_{i}^{(t)}, x^{(t+1)}_j\right) 
	    & +\frac{1}{n_2^2} \sum_{i, j=1}^{n_2} \mathcal{K}\left(x^{(t+1)}_i, x^{(t+1)}_j\right),
    \end{aligned}
\end{equation}
where $\mathcal{F}$ is Hilbert space, $\mathcal{K}(\cdot)$ is the Gaussian kernel function and $n_1, n_2$ are the number of samples drawn from $\mathbf{X}^{(t)}$ and $\mathbf{X}^{(t+1)}$. If $\mathbf{X}^{(t)}$ = $\mathbf{X}^{(t+1)}$, the MMD value is 0. A large MMD indicates a substantial data shift, necessitating the consideration of a large update size for stable recommendation quality. Given that MMD is non-negative, we simply define the following equation to determine the compression ratio:
\begin{equation}
r = \lceil \frac{1}{C\left(2\sigma(\mathrm{MMD}(\mathbf{X}^{(t)},\mathbf{X}^{(t+1)}))-1\right)} \rceil,
\end{equation}
where $\sigma$ denotes sigmoid function and $C\in(0,1]$ is a constant to further scale the calculated value to a proper range. In our experiments, we find $C=0.2$ a universally applicable value for different datasets.

% \begin{table}[h]%[tb]
% 	\renewcommand\arraystretch{1.0}
% 	% 	\caption{Dataset Statistics}
% 	\begin{center}
% 		\begin{tabular}{ccccc}
% 			\hline
% 			Dataset & Tmall & Xing  \\ \hline
% 			% clicks & 557,248 & 8,326,407 & 982,961  \\
% 			training sessions & 351,268 &  78,276\\
% 			test sessions & 25,898 &  11,315\\
% 			\# of items & 40,728 & 58,035  \\
% 			average lengths & 6.69 & 5.24  \\
% 			\hline
% 		\end{tabular}
% 	\end{center}
% 	\caption{Dataset Statistics.}
% 	\label{Table.1}
% \end{table}
\begin{table}[ht]%[tb]

		\begin{center}
			\resizebox{.45\textwidth}{!}{
			\begin{tabular}{c|c|c|c|c|c|c}
				\hline
				Dataset& $D_1$ & $D_2$ & $D_3$  & $D_4$ & $D_5$ & test   \\ \hline	
				Gowalla & 118,858 & 305,543 & 581,928 & 914,910 & 1,261,076 & 99,711 \\
				Lastfm   &133,948   &252,333  & 648,666  & 1,267,288  & 2,538,670   &607,274  \\
				\hline
			\end{tabular}}
		\end{center}
		\caption{Dataset Statistics}	
		\label{Table.1}
		\vspace{-2em}
	\end{table}

\subsection{Training Scheme}
The proposed compositional codes-based compression framework is model agnostic, which means it can apply to a wide range of session-based recommendation models with distinctive architectures in a plug-and-play fashion. Given the session embedding $\bm{s}$ obtained by any session-based recommendation model and the original item embedding $\textbf{X}$, the cloud-side model is trained with the cross-entropy loss defined as:
\begin{equation}
	\mathcal{L}_{rec}=-\sum_{s}\sum_{v\in s} \mathbf{y}_{sv} \log \left(\hat{\mathbf{y}}_{sv}\right)+\left(1-\mathbf{y}_{sv}\right) \log \left(1-\hat{\mathbf{y}}_{sv}\right),
\end{equation}
where $y_{sv}$ is the ground truth and $\hat{y}_{sv}=\bm{X}_{v}^{\top}\bm{s}$ is the predicted possibility. After the cloud-side model is well trained, we duplicate all the parameters of it except the item embedding table $\bm{X}$, and use them to initialize the on-device model. To train the one-hot code matrices and codebooks for model compression, we minimize the mean squared error (MSE) loss which calculates the difference between the original embedding (fixed) and the reconstituted embedding:
\begin{equation}
\mathcal{L}_{mse} = \underset{\bm{O}, \bm{E}}{\operatorname{argmin}} \left\| \bm{O}\bm{E}-\mathbf{X}\right\|^{2}.
\end{equation}
After the deployment, every time when the update is required, the above process would be repeated. But it should be noted that the codebooks are only partially retrained for update compression.

% \subsection{Session-Based Recommendation}
% Early session-based recommendation models \cite{shani2005mdp,rendle2010factorizing, yin2016spatio} were based on Markov Chain and primarily focused on modeling the temporal order between items. With the advent of deep learning, networks such as RNNs \cite{hidasi2015session, chen2020sequence,zhang2018discrete} have been modified to model session data and handle the temporal shifts within sessions. Subsequent models, such as \cite{li2017neural} and STAMP \cite{liu2018stamp}, employ attention mechanisms to assign different priorities to items when profiling a user's main interests. Graph-based methods \cite{wu2019session} design various kinds of graph structures to model session data. For example, SR-GNN \cite{wu2019session} constructs session graphs for every session and designs a gated graph neural network to aggregate temporal information and item relations into session representations. Xia \textit{et al.} \cite{xia2021aaai, xia2021cikm} proposed to integrate self-supervised learning into session-based recommendation to boost performance. These models are successful when generating accurate recommendations but they are too large to run on resource-constrained devices.

\section{Experiments}
\subsection{Experimental Settings}
\subsubsection{Datasets.}
We evaluate our model on two real-world benchmark datasets: \textit{Gowalla}\footnote{https://snap.stanford.edu/data/loc-gowalla.html} and \textit{Lastfm}\footnote{http://mtg.upf.edu/static/datasets/last.fm/lastfm-dataset-1K.tar.gz}.
% and \textit{Diginetica}\footnote{http://cikm2016.cs.iupui.edu/cikm-cup/}. 
\textit{Gowalla} is a location-based check-in dataset to do next-POI recommendation. We filter out sessions that are shorter than 2 and longer than 300 and the item number is 37,722.  
\textit{Lastfm} is used for music artist recommendation and we treat user's transactions in 8 hours as a session. We keep the top 10,000 most popular artists and filter out sessions that are longer than 50 or shorter than 2 items. For both datasets, we split them into five training sets $D_1, D_2, D_3, D_4, D_5$ and one test set based on temporal order. The proportion of the five training sets for \textit{Gowalla} is 1:3:6:10:15 and 1:2:5:10:20 for \textit{Lastfm}. Since in recommendation scenarios, we consider that the recommendation model will not update frequently in the late periods because the model already has good capability, we make the proportion of the five training sets progressively increase. $D_5$ set contains all training sessions. To augment and label the training and test datasets, we follow the convention to adopt a sequence splitting method, which generates multiple labeled sequences with the corresponding labels $([v_{s,1}], v_{s,2}), ([v_{s,1},v_{s,2}], v_{s,3}), ...,([v_{s,1}, v_{s,2}, ..., v_{s,l-1}], v_{s,l})$ for every session $s = [v_{s,1}, v_{s,2}, v_{s,3}, ..., v_{s,l}]$. Note that the label of each sequence is the last consumed item in it. The number of sequences of split datasets are presented in Table \ref{Table.1}.

\subsubsection{Base Models.}
We apply the proposed framework to the following representative session-based recommendation methods which are with distinctive architectures to validate its effectiveness and efficiency for model and update compression:
\begin{itemize}[leftmargin=*]

	\item \textbf{NARM} \cite{li2017neural} is an RNN-based model which employs an attention mechanism to capture users' main purpose and combines it with the temporal information to generate recommendations.
	\item \textbf{SR-GNN} \cite{wu2019session} is a gated GNN-based model that employs a soft-attention mechanism to compute the session embeddings.
        \item {\textbf{SASRec} \cite{kang2018self} is a Transformer-based sequential recommendation model to capture long-term semantics and predict based on relatively few actions.}
\end{itemize}

\subsubsection{Metrics}
We use Pre@K (Precision) and NDCG@K (Normalized Discounted Cumulative Gain) to evaluate the recommendation results where K is 5 or 10. Precision measures the ratio of hit items and NDCG assesses the ranking quality of the recommendation list. 

\subsubsection{Hyperparameter Settings}
As for the setting of the general hyperparameters, we set the mini-batch size to 100, the $L_2$ regularization to $10^{-5}$, and the embedding dimension to 128 for both datasets. All the learnable parameters are initialized with the Uniform Distribution $U(-0.1,0.1)$ for NARM, SASRec and Normal Distribution $N(0,0.1)$ for SR-GNN. or Normal Distribution. For NARM, we set the number of GRU layer as 1 and dropout rate as 0.2 for both datasets. For SR-GNN, we set the number of GNN layer as 1. For SASRec, we use one Transformer layer, one attention head and one attention block and dropout rate as 0.2 on both datasets. We optimize the model using Adam with a learning rate of 0.001. We empirically set the temperature $\epsilon$ in Gumbel-Softmax to 0.2.

% \subsection{Effectiveness of Baselines on the Cloud}
% The cloud models keep retraining when accumulating larger datasets. Therefore, it is crucial to continually evaluate their performance and improvement with each new training data. We report their performances in Table \ref{Table.2}. These results highlight that all our baseline models exhibit improvement with the accumulation of new training data. Notably, the Transformer-based sequential model, SASRec, excels on both datasets, underlining its superior capability in capturing user interests in sequential tasks. Our second most effective baseline is the NARM model, which outperforms the graph-based model SR-GNN. Despite SR-GNN's exceptional performance in other session-based recommendation benchmarks, it has demonstrated limitations in learning sequential representations and dependencies in Gowalla. It's also important to clarify that the objective of this paper is not merely to maximize the recommendation accuracy of the model, but rather to focus on updating the on-device models efficiently and cost-effectively.

\begin{table}
\footnotesize
\caption{On-device performance at $D_1$ on Gowalla.}
  \vspace{-1em}
\label{table:2}
\centering
\begin{tabular}{ccccccc} 
	\hline
	Model & n & CR & Prec@5 & NDCG@5  & Prec@10  & NDCG@10 \\\hline
				  &  10   & 12  & 8.01 & 5.86  & 10.57 &  6.70  \\
			NARM&  20   & 6  & 8.78 &  6.49 & 11.42 &  7.35  \\
				&  40   &  3 & 9.33  & 6.94  & 12.19  &  7.84  \\ \hline
                 NARM-C  &  --   & -- & 9.32  & 6.97  &  12.24   & 7.92    \\\hline

				   &  10   & 12  & 11.81 &  7.76 & 13.14 & 8.34   \\
				SR-GNN  & 20 &  6 & 15.11 & 9.66  & 17.94 & 10.34   \\
				    &  40   &  3 & 15.24 & 9.75  & 18.25 &  10.57  \\ \hline
                SR-GNN-C  &  --   &  -- &  15.27 & 9.55  &  18.85   &  10.20   \\\hline

                         &   10  &  12 & 10.63 &  8.32 & 13.30 & 9.18   \\
				  SASRec &  20   &  6 &  15.07&  12.16 &17.92  & 13.05   \\
				    &   40  & 3  & 16.36 & 13.30  & 19.21 & 14.21  \\ \hline
                     SASRec-C  &  --  &  -- & 15.99  &  12.33 &  19.17   & 13.37    \\\hline

	\end{tabular}
  \vspace{-1em}
\end{table}

\begin{table}
\footnotesize
\caption{On-device performances at $D_1$ on Lastfm.}
  \vspace{-1em}
\label{table:3}
\centering
\begin{tabular}{ccccccc} 
	\hline
	Model & n & CR & Prec@5 & NDCG@5  & Prec@10  & NDCG@10 \\\hline
				  &  10   &  9 & 4.67 & 3.05  & 7.17 & 3.86   \\
				NARM &20 & 5 & 4.66 & 3.07  & 7.14 & 3.87  \\
				&  40 & 2    & 4.74 & 3.10  & 7.25 & 3.91   \\    \hline
                   NARM-C  &  --   &  -- & 4.74 & 3.11  & 7.22   &3.91      \\\hline

				   &  10   &  9 & 4.21 & 2.87 & 6.26 & 3.53   \\
				SR-GNN  &20 & 5 & 4.51 & 3.12  &6.54 & 3.77   \\
				    &  40   & 2 & 4.56 & 3.18 & 6.58 & 3.82   \\ \hline
                     SR-GNN-C  &  --   &  -- & 5.86 &  4.00 &  8.63  & 4.89     \\\hline

                         &   10  & 9  & 4.43 & 2.97 & 6.65 &  3.69  \\
				  	SASRec & 20    &  5 & 4.82 & 3.36  & 7.00 &  4.05  \\
				    &  40   & 2  & 4.72 & 3.30 & 6.88 & 3.99  \\ \hline
                     SASRec-C  &  --  &  -- &4.80  & 3.38  &  6.94  &4.07      \\\hline

	\end{tabular}
 \vspace{-1em}
\end{table}

\begin{figure*}[t]  
    \centering
	\vspace{-10pt}
    \includegraphics[width=\textwidth]{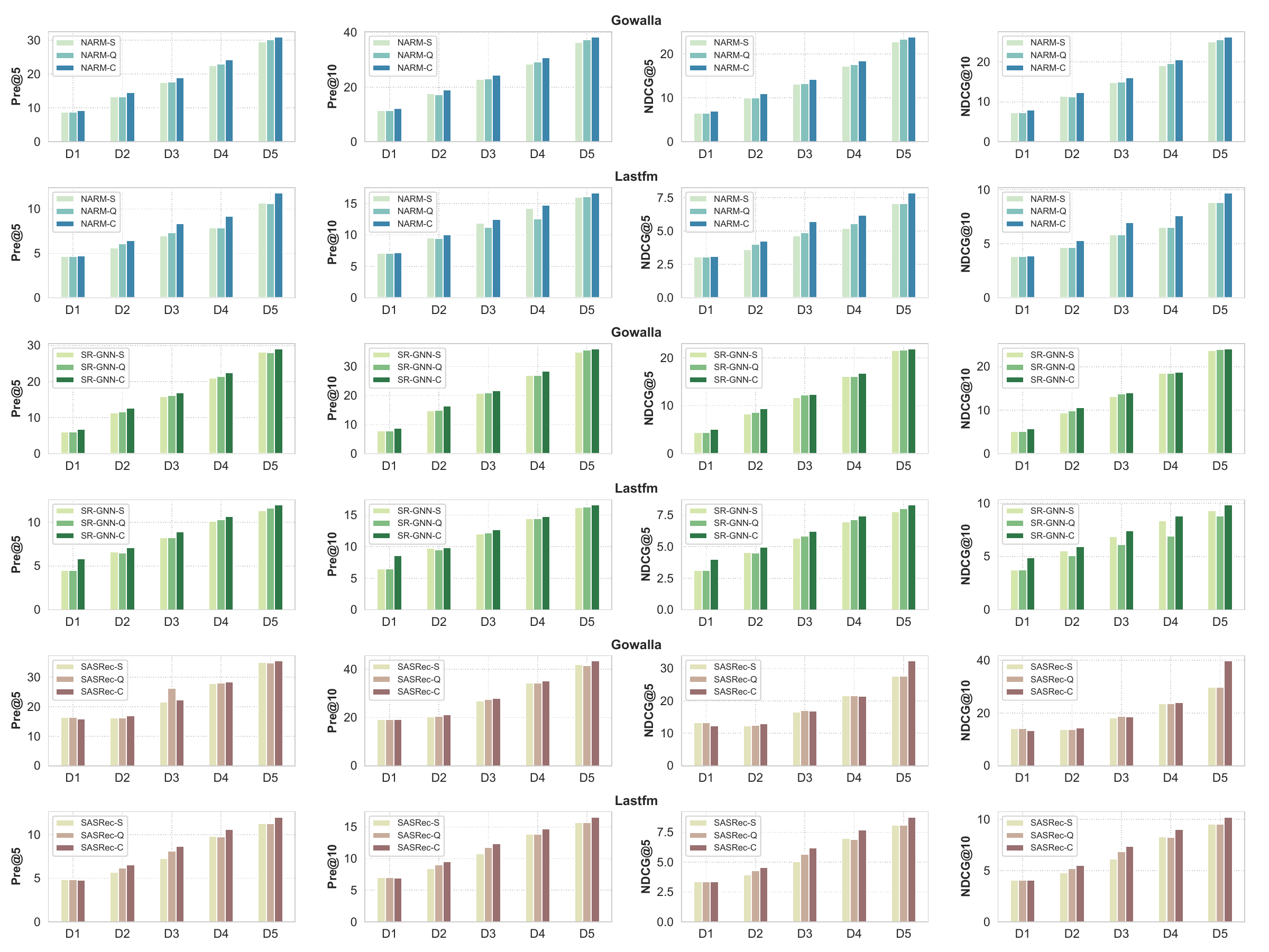}%
    \caption{Performance comparison using stack- and queue-based update compression.}
    \label{figure.4}
\end{figure*}

\begin{figure*}[t]
	\centering
	\includegraphics[width=\textwidth]{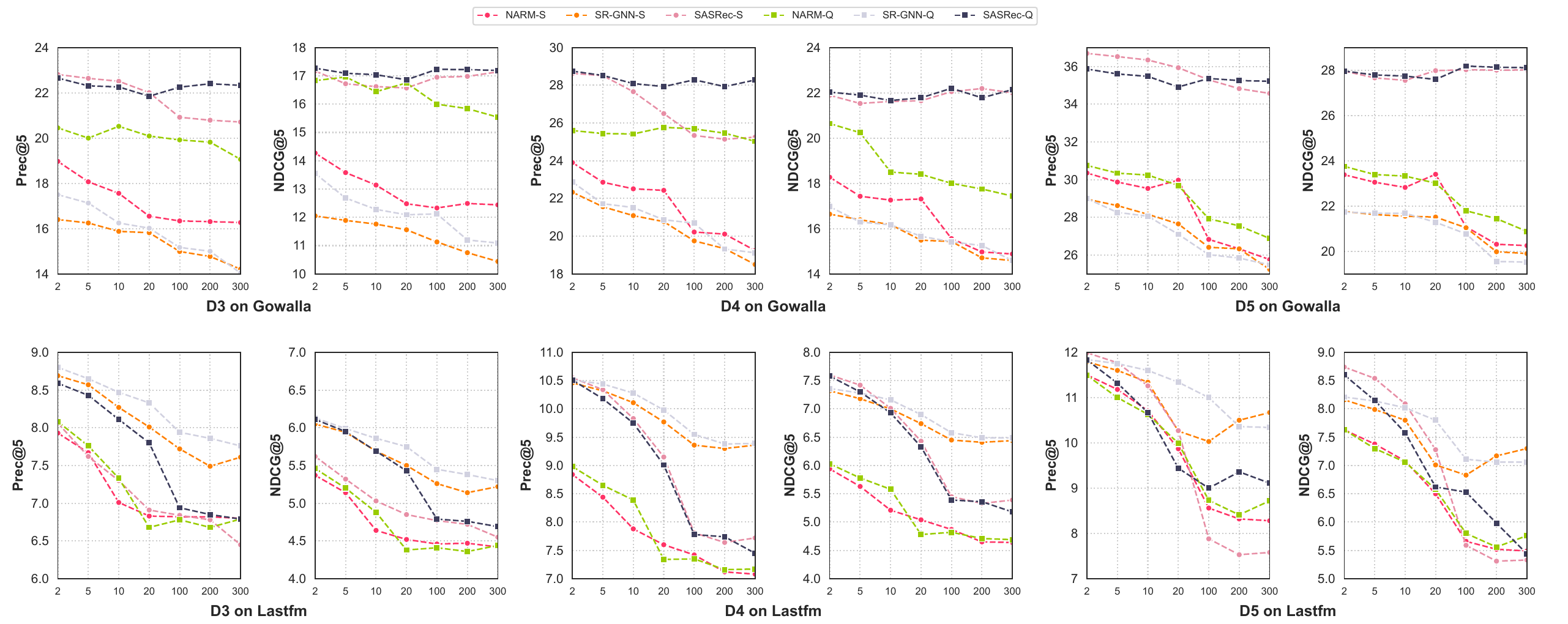}
	\caption{Sensitivity analysis.}
	\label{figure.5}
\end{figure*}

\subsection{Effectiveness of Model Compression}
In order to validate the efficacy of compositional codes, we first evaluate the recommendation performances of the compressed models at various compression ratios. In this experiment, the proposed model update approaches are not involved and the results are obtained on data slice $D_1$. Recall that, $n$ represents the number of codebooks, and $k$ stands for the number of vectors in each codebook. Due to space limitations, we only showcase the performances under different values of $n$ in Table \ref{table:2} and Table \ref{table:3} and here we set $k=32$ ($n$ is also more decisive for the compression ratio according to Eq. \ref{eq: cr}). CR in the table headers stands for compression ration and -C represents the uncompressed cloud model. We observe that with the decrease of $n$, the compression ratio increases while the performance declines, which is logical that extreme compression leads to inferior performance. Besides, the three models are more sensitive on Gowalla compared to Lastfm. Compared with NARM and SR-GNN, SASRec appears to be more sensitive to the compression ratio on Gowalla and displays comparable performance with its uncompressed counterpart only when the compression ratio is minimal. Based on these findings, we have selected suitable $n$ values for the three models across the two datasets. In our experimental setup, we choose $n$=20 for NARM and SR-GNN, $n$=40 for SASRec on both Lastfm and Gowalla. These selected values will be utilized for the ensuing experiments.

\subsection{Efficacy of Update Compression}
In this part, we evaluate the efficacy of the proposed stack-based and queue-based approaches. Due to space limitations, we only showcase model performances when the update size is reduced to 10\% of the size of the compressed model, as illustrated in Figure \ref{figure.4}. \textbf{Method-C} means the corresponding cloud model, \textbf{Method-S} and \textbf{Method-Q} represent device models that are using stack-based and queue-based approaches to do update compression, respectively.
Several observations can be made from Figure \ref{figure.4}:
\begin{itemize}[leftmargin=*]
\item With more data added, the model performances consistently improve. On a general note, our proposed stack-based and queue-based approaches both deliver comparable performances when compressed to 10\% for all the three base models across different data slices. Specifically, the 10x update compression only leads to an accuracy loss within 10\%. This suggests that our approach to compressing updates is not only effective but also universally applicable to different types of session-based recommendation models and datasets. According to Eq. (\ref{eq: tcr}), when the model compression size is considered, for SR-GNN and NARM, the update size is 60x smaller than the item embedding table in the cloud. Note that the model compression and update compression methods can be easily implemented on other models.
% \item Of the three models, SASRec consistently outperforms the others across most data splits, highlighting its proficiency in modeling user intents in session-based contexts. Conversely, SR-GNN outperforms NARM on both datasets, indicating the superiority of graph structure on modeling user intents.
\item Comparing the stack and queue approaches, the latter generally delivers marginally superior results on both datasets. This could be attributed to the queue approach's ability to consistently maintain the most recent parameters, while the stack approach focuses on the addition of new information. Moreover, on $D_1$, the two approaches have the same performance because there is no model update in this period. And on $D_2$, the two have very close performances because it is the first time to compress updates and they both recycle the same amount of vectors from $D_1$. 
\end{itemize}

\subsection{Sensitivity Analysis}
To investigate the correlation between update compression ratio and model accuracy, we select representative update compression ratio values \{2, 5, 10, 20, 100, 200, 300\} and conduct experiments with them. The results are displayed in Figure \ref{figure.5}. The top row of the figure illustrates the results on the Gowalla dataset, while the bottom row depicts the results on Lastfm. Given that performances of stack-based and queue-based approaches on $D_2$ are relatively similar, we only perform this experiment on the last three data slices.
We construct six models using NARM, SR-GNN, and SASRec, each having two variants employing either the stack or the queue approach. These models are referred to as $\textbf{NARM-S}$, $\textbf{NARM-Q}$, $\textbf{SR-GNN-S}$, $\textbf{SR-GNN-Q}$, $\textbf{SASRec-S}$, and $\textbf{SASRec-Q}$. Observations drawn from Figure \ref{figure.5} include:

\begin{itemize}[leftmargin=*]
    \item In general, with the increase in compression ratio, the performances of all six models steadily decline on both datasets.
    \item Nearly all models exhibit low sensitivity on Lastfm when the compression ratio is large, such as at 100, 200, and 300. This is because the maximum number of total learnable vectors is $nk=1280$, and the number of learnable vectors are very small and close after compression. Conversely, when the compression ratio is smaller, there is a larger gap between model performances. On Gowalla, models are more sensitive, further showing that learning user interests on it is heavily dependant on parameter size.
    
    \item On Gowalla, SASRec-Q marginally surpasses SASRec-S in most cases, suggesting that the queue approach is more effective. On Lastfm, the best performance is achieved by SR-GNN-S, and we can also observe that some stack-based approaches outperform queue-based approach, indicating that stack-based approach has its unique advantages on certain datasets. The decision between stack and queue-based approaches should be contingent upon the specific dataset in use.
\end{itemize}
\vspace{-10pt}

% \begin{figure*}[t]
% 	\centering
% 	\includegraphics[width=\textwidth]{figure4_new.pdf}
% 	\caption{Sensitivity analysis.}
% 	\label{figure.6}
% \end{figure*}

% \begin{figure}[t]
% 	\centering
% 	\includegraphics[width=0.5\textwidth]{figure5.pdf}
% 	\caption{Performances of on-device models when CR=10 using stack.}
% 	\label{figure.5}
% \end{figure}
\begin{figure}[t]
	\centering
	\includegraphics[width=0.5\textwidth]{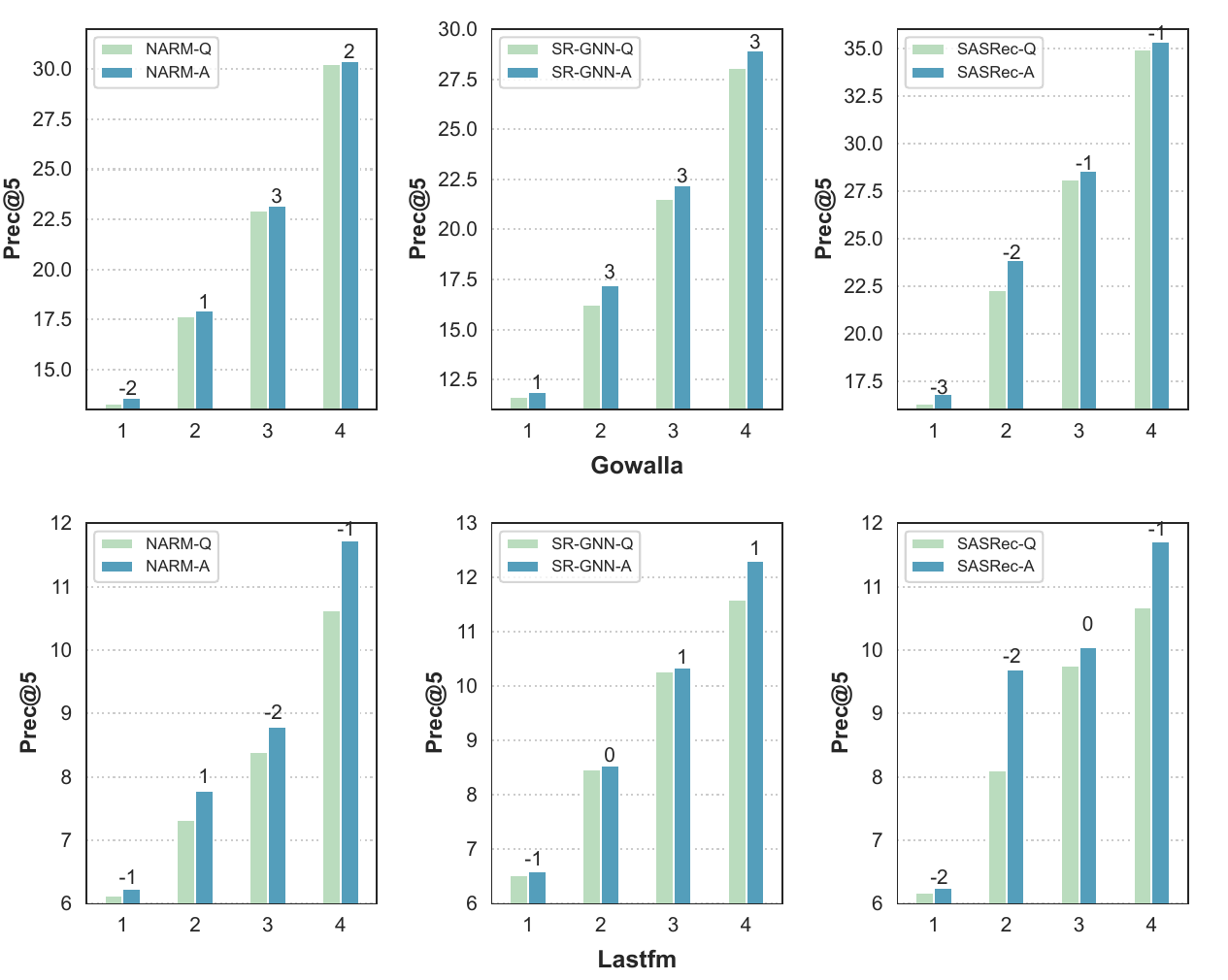}
	\caption{Self-adaptive update compression performances.}
	\label{figure.6}
\end{figure}
\subsection{Adaptiveness Analysis}
To examine the effectiveness of our proposed self-adaptive update compression method, we apply this technique to both datasets. Initially, we utilize Equations (16-17) to compute the Maximum Mean Discrepancy (MMD) value between each pair of adjacent data slices and their corresponding update compression ratios. Subsequently, we derive the self-adaptive results and display the Prec@5 results in Figure \ref{figure.6}. Given that the queue-based approach slightly surpasses the stack-based approach, we directly compare self-adaptive update compression with queue-based update compression. In this context, \textbf{NARM-A}, \textbf{SR-GNN-A} and \textbf{SASRec-A} denote their models with the self-adaptive update compression. The figure exhibits the difference values of compression ratios for the two methods (negative values indicate the self-adaptive compression ratio is smaller than 10). The values under the X-axis represent each update compression in chronological order. From these results, we consider that the adaptive update compression proves beneficial in achieving optimal performance in session-based scenarios. There are instances where minor additional compression does not significantly degrade accuracy. Moreover, we observe that employing smaller compression ratios in the initial period and larger ones later yields superior performances on both datasets. Therefore, adjusting update compression ratios based on the cloud model's item embedding emerges a strategic approach to achieve a trade-off between performance and efficiency.

% \begin{table}[t]%[tb]
% \footnotesize
	
% 	\begin{center}
% 	{
% 		\begin{tabular}{ccccc|cccc}
% 	        \toprule
%             \multicolumn{5}{c}{Gowalla} & \multicolumn{4}{c}{Lastfm} \\
% 			\hline
% 			Method& $D_{1-2}$ & $D_{2-3}$ & $D_{3-4}$ & $D_{4-5}$ & $D_{1-2}$ & $D_{2-3}$ & $D_{3-4}$ & $D_{4-5}$  \\ \hline
% 			 NARM & 8 & 11 & 13 & 12 & 11  & 11  & 8  &9 \\
% 			SR-GNN  & 11 & 13 & 13 & 13 &  9 &  10 & 11  & 11\\
% 			 SASRec & 7 & 8 & 9 & 9 &  8 & 8  &  10 & 9 \\
% 			\hline
% 		\end{tabular}}
% 	\end{center}
% 	\caption{Adaptive update compression ratios.}
%  \label{Table:4}
% 	\vspace{-10pt}
% \end{table}

\section{Conclusion}
On-device recommendation is becoming popular recently since the paradigm of on-device machine learning enables large models to run on resource-constrained devices and meanwhile provide secure services. But one issue is that how to efficiently update on-device models to meet the need of model retraining. Unlike cloud-based models, device models can not retrain itself due to limited resources. In this paper, we address this issue by proposing an efficient method to compress the model updates and then transfer them to the device to finish retraining where such a method can compress the updates under the bandwidth limit with comparable performances and the compressed updates should be easily applied for device models to update. Extensive experimental results of three different models on two benchmarks demonstrate the effectiveness and efficiency of our proposed method.
\section{Acknowledgement}
This work is supported by Australian Research Council Future Fellowship (Grant No.FT210100624), Discovery Project (Grant No. DP190101985), and Industrial Transformation Training Centre (Grant No.IC200100022).  
\bibliographystyle{ACM-Reference-Format}
\bibliography{ref}
\end{document}